\documentstyle[preprint,aps]{revtex}
\begin{document}

\title{Anomalous magnetic properties of Rh$_{13}$ clusters}
\author{Yang Jinlong$^{1,2}$, F.Toigo$^2$, Wang Kelin$^1$,
         and Zhang Manhong$^1$}
\address{$^1$Center for Fundamental Physics,University of
Science and Technology of China,\\
Hefei, Anhui 230026, P.R. China}
\address{$^2$Dipartimento di Fisica dell'Universit\`a
and Consorzio Interuniversitario di Struttura della Materia INFM,
via Marzolo 8, 35100 Padova, Italy}

\date{}
\maketitle
\begin{abstract}

Electronic structures of 13-atom Rh clusters with three possible
high-symmetry geometries are studied using the discrete-variational
local-spin-density-functional method. The ground state is found to
be the icosahedral structure, and a total magnetic moment of 15$\mu_B$
is obtained for the cluster. This value is anomalously smaller than those
for clusters with lower symmetries, but in agreement with recent experiments.
The magnetic interactions between the central and surface atoms of the cluster
are {\it not} fully ferromagnetic, and a small amount of antiferromagnetic
interactions is found to be mixed in. An energy parameter
is introduced to explain the anomalous magnetic properties,
which is found to be also
useful for judging whether some techniques can or must be used in the
local-spin-density-functional calculations.
\end{abstract}


\narrowtext

Transition-metal (TM) clusters have been the subject of widespread
investigations
in recent years because of their promising practical applications in
developing new magnetic materials with large moments.\cite{chicago,yang}
As it is well-known, all $3d$, $4d$ and $5d$ TM atoms
have a finite magnetic moment due to the Hund's-rule
coupling in their unfilled $d$ shells, and only $3d$ Fe, Co and Ni atoms are
able to retain these moments at a much reduced level in the bulk environment.
On the other hand, small TM clusters, as a new state of materials,
may have magnetic properties different from their bulk phase and atoms.
For free Fe, Co and Ni clusters, theoretical
calculations\cite{dunlap,khanna,li} and
experimental measurements\cite{deheer,bucher,louderback}
have shown they have magnetic moments per atom that are bigger
than the corresponding bulk values. For the
other TM clusters, however, we know a little
about their magnetic properties. Sometimes the conclusions from experiments
contradict predictions from theories. Theoretical calculations\cite{liu,pastor}
usually predict
large moments for small clusters, while experimental
measurements\cite{douglass}
give nonmagnetic results in the experimental resolution limits.
Whether these clusters can be magnetic was a question until
recently, when Cox {\it et al.}\cite{cox}
observed experimentally that clusters
of $4d$ nonmagnetic solid Rh exhibit a permanent magnetic moment.
This moment can be as large as 1.1$\mu_B$/atom.
This experiment confirmed the theoretical prediction by Reddy
{\it et al.}\cite{reddy} that 13-atom
clusters of $4d$ Pd, Rh and Ru
all have nonzero magnetic moments. Carefully comparing the
results of Cox {\it et al.} and of Reddy {\it et al.}, however,
one can find
a quantitative discrepancy between experiment and
theory.\cite{falicov}
The total moment
of Rh$_{13}$ was measured to be 11.5$\mu_B$, just above half
the 21$\mu_B$ predicted by theory.
Moreover, theory gives a full
ferromagnetic (FM)-interaction
picture for the cluster, while experiment suggests a more complicated
picture in which not all spins are parallel.

Motivated by the discrepancy mentioned above, we have performed a
first-principles study on the electronic structures of Rh$_{13}$ clusters
with three possible high-symmetry geometries. Our results have removed
the discrepancy between experiment and theory, and are in agreement with
the experimental measurement.
Furthermore, we found
an anomalous relationship between the cluster symmetry and the magnetism
of Rh$_{13}$ clusters, i.e. the total moment of the icosahedral
Rh$_{13}$ cluster is smaller than that of the lower-symmetry clusters in a
wide range of interatomic spacings.
This is particularly remarkable because it is always believed that the
magnetic moment of a cluster is a consequence of the reduced dimensionality
and increased symmetry. We will rationalize this anomalous relationship
in terms of an energy parameter which turns out to be
useful also for judging whether some techniques can or must be used
in a local-spin-density-functional(LSD) calculation.

The three possible high symmetries we chose for Rh$_{13}$ clusters are
I$_h$, O$_h$ and D$_{3h}$ respectively. The I$_h$ point group, being
that of an
icosahedron, is too highly symmetric for any crystal.
The O$_h$ structure is a cuboctahedron, which is a compact portion of
the fcc crystal lattice. The D$_{3h}$ structure is obtained from the O$_h$
cluster by rotating any triad of nearest-neighbor surface atoms by
60$^\circ$ about their center. This third cluster is a compact portion
of the hcp lattice.

The binding energy and electronic structure of clusters
were calculated using the discrete-variational (DV) LSD
method. It is a kind of molecular orbital calculation method
and its theoretical foundation is LSD theory.
Since it has been described in detail elsewhere,\cite{delley1,delley2}
we do not give a further description here.

There are two computational schemes within the DV method. In the
first one, the exact cluster charge density $\varrho({\bf r})$ is
replaced approximately by a model density, which is a superposition
of radial densities
centered on cluster atoms via diagonal-weighted Mulliken
populations.\cite{rosen} In the second, a multipolar, multicenter model
density is used to fit $\varrho({\bf r})$ with a least-squares
error-minimization procedure.\cite{delley1} One of the methods used by
Reddy {\it et al.} is the DV method: unfortunately they did not
specify which scheme they used. In our calculations
we adopted the second scheme
which leads to a true self-consistent solution and therefore to the
more precise results compatible with the method.
Our calculations differ from those of Reddy {\it et al.},
in two important respects: $(a)$ we expanded the basis
set to include Rh $5p$ orbital and roughly optimized the Rh
$4d^{8+x}5s^{1-(x+y)}5p^y$ ($0<x<1$ and $0<x+y<1$) configurations for
the atomic basic functions in order to minimize the calculated cluster energy.
The optimal basis set was found to be the numerical atomic basic functions of
the Rh $4d^85s^{0.9}5p^{0.1}$ configuration, and $(b)$ we dropped the
Lorentzian
broadening for determining the occupation number near the Fermi energy (E$_F$).
This ensures that our results are
the real solutions of the Kohn-Sham equations.
The price to be paid is a slower convergence.

For each electronic structure calculation,
we used several input potentials and started the calculations from
configurations with various magnetic moments.

For most cases, we obtained
only one self-consistent solution. In certain interatomic spacings,
however, more than one self-consistent solution can exist. These solutions
correspond to local minima of the cluster energy as a function of the
cluster moment. For these cases, the one which gives the largest
cluster binding energy was chosen as our final solution.
In addition, we used the two different forms of the
exchange-correlation potential proposed by von Barth and Hedin\cite{barth} and
by Perdew and Zunger.\cite{perdew}
The calculated results are
found to be independent on the form of the exchange-correlation potential.

The binding-energy curves versus the distance $r$ between the central
and surface atoms are plotted in Fig. 1 for all the clusters. From Fig. 1, we
can determine the equilibrium configuration of a cluster, as presented
in Table I. The ground state is found to correspond to the I$_h$ cluster,
which is
more stable than the D$_{3h}$ and  O$_h$ clusters by 0.45 eV
and 1.35 eV respectively. In the I$_h$ cluster, the binding energy per atom
is 4.01 eV,
about 30\% smaller than the bulk cohesive energy which is 5.75 eV. Compared
to the bulk interatomic spacing of 5.1 a.u., one may find
small bondlength contractions ($<5\%$) in all clusters. Such a contraction
effect was observed by extended x-ray-absorption fine structure measurements
in Cu and
Ni clusters and the contraction ratio was found to be proportional to the
surface-to-volume ratio of the cluster,\cite{apai}
so it is believed to be
a consequence of surface effects. Table I also lists the results of Reddy
{\it et al.}.\cite{reddy} The bondlengths
in the two calculations are almost same, but the
binding energies have large differences. We believe that the smaller binding
energies of Reddy {\it et al.} are a result of the smaller basis set
used.

Figure 2 presents the cluster moment as a function of $r$ for the
three clusters.
The O$_h$ and D$_{3h}$ clusters carry the same moment
(19$\mu_B$), which remains unaltered over the range of $r$ spanned
in this figure.
O$_h$ and D$_{3h}$ clusters
are expected to exhibit similar magnetic properties
since each surface atom of the two clusters sees an identical
nearest-neighbor environment.
The total moment of the I$_h$ cluster increases from 15$\mu_B$ to
17$\mu_B$ to 21$\mu_B$ with the increase of $r$. In a wide range of
$r$ ($r<$5.0 a.u.), the total moment of the I$_h$ cluster is smaller
than that of
the O$_h$ and D$_{3h}$ clusters. This result obviously contradicts the
rule\cite {dunlap} for clusters of iron-group atoms occupying
equivalent volumes:
the higher the order of the group, the larger the cluster moment.
{}From Fig. 2, we see that the I$_h$ cluster does have the largest total
moment when $r>$5.0 a.u., and obeys the above rule.

Why does the I$_h$ cluster reduce rapidly its moment with the decrease
of $r$ while  O$_h$ and D$_{3h}$ clusters do not? We found the answer by
analyzing the one-electron energy levels around E$_F$.
First, we define an energy parameter $\Delta$E.

For a cluster whose highest occupied molecular orbital (HOMO) is
partially occupied, $\Delta$E is the
energy difference between the HOMO and its  closest-in-energy spin-opposite
molecular orbital (CSMO) which can be either occupied or unoccupied.

If the HOMO is fully occupied, then
$\Delta$E is either the energy difference between the
HOMO and its unoccupied CSMO or  between the lowest unoccupied molecular
orbital (LUMO) and its occupied CSMO, depending on which one is the smaller.
In our calculations, we always found partially occupied HOMO's.
Generally speaking, the order of the HOMO (or LUMO) and its CSMO in a
cluster
can be altered by changing interatomic spacings, if the value
of $\Delta$E is small, and this will result in the change of the cluster
magnetic
moment. The I$_h$ cluster is a case. We found that $\Delta$E in this cluster
is very small (e.g., $\Delta$E$\approx$0.05 eV for $r$ near its
equilibrium value). For
O$_h$ and D$_{3h}$ clusters, however, the HOMO's are far from their
CSMO's in energy and the values of $\Delta$E are about 0.5 eV. So, it
is not easy to alter the order of the HOMO and its CSMO by simply changing
the interatomic spacings of these clusters. This is the reason why
the moments of the O$_h$ and D$_{3h}$ clusters remain unaltered
over the range of $r$ spanned in Fig. 2.

Table II lists the total magnetic moments of all the clusters at their
equilibrium configurations. The moment per atom of the I$_h$ cluster is
calculated to be 1.15$\mu_B$, which is in good agreement with the experimental
one (0.88$\pm$0.16$\mu_B$).\cite{cox} Compared to our total moment
of 15$\mu_B$, Reddy {\it et al.}\cite{reddy}
obtained a larger moment of 21$\mu_B$. The discrepancy, we believe,
arises due to our modifications in the calculations. Since the effect of
enlarging the basis set is obvious, we do not discuss it and focus our
attention on other points. In their calculations, Reddy {\it et al.}
used a 0.05 eV Lorentzian broadening to determine the occupation number
near E$_F$. However, we have seen above that in the I$_h$ cluster
$\Delta$E is also about 0.05 eV.
Since the broadening parameter is of the same order of magnitude of
$\Delta$E, the occupation numbers are affected by its value.
We note that the broadening technique is used to accelerate the iteration
convergence in most LSD calculations.\cite{dunlap,li,liu,pastor}
Our result indicates that one must be very careful in choosing the
value of the broadening parameter when $\Delta$E is small.

It is well known that the Kohn-Sham equations in the local-density-functional
(LDF) scheme have a unique solution for a given system. In the LSD scheme,
however, solving the equations --- simultaneously optimizing the spin of the
system --- can yield more than one solution. This is to say, in LSD
calculations the
self-consistent solution may depend on the input potential.
Actually, we found two self-consistent solutions for the
I$_h$ cluster with $r$=4.84 a.u. :
the total moment of the first is 15$\mu_B$ while in the second it is
21$\mu_B$. The two solutions correspond to
two local minima of the cluster energy as a function of the
cluster moment. The cluster binding energy of the former was calculated
to be larger than that of the latter by 0.35 eV. So, the solution we discussed
above is the global minimum while the solution obtained by Reddy
{\it et al.} is only a local minimum.

When should one look for multiple
solutions in a LSD calculation? Again, we link the answer to our energy
parameter $\Delta$E. We suggest that when one finds $\Delta$E to be small,
say less than 0.1 eV, one should consider
the possibility of multiple solutions in the calculation.

The local magnetic moments of the three clusters at their equilibrium
configurations are also shown in Table II. For all three clusters, the
local moment of the central atom is smaller than that of surface atoms.
This observation agrees with results for clusters of
iron-group atoms.\cite{yang,li}
{}From Table II, we found a complicated magnetic-interaction picture for
all three clusters, in agreement with the experimental suggestion.\cite{cox}
The magnetic interactions between the central and
surface atoms are mainly FM, but a small amount of antiferromagnetic (AFM)
interactions is found to be mixed in: the local moments of the central Rh
$5s$ and $5p$ align in an opposite direction to those of the central Rh $4d$
and
surface atoms. Such a small amount of AFM interactions can not be neglected
because it may affect the temperature and external magnetic field
dependences of the magnetic properties of the cluster.
To substantiate our magnetic-interaction picture, we have prepared
the spin-density distribution plots
on two typical planes of the icosahedron (Fig. 3). In Fig. 3, a small
amount of the negative polarizations is apparent.

Figures 4(a) and 4(b) show the densities of states (DOS) for the
majority- and minority-spin electrons in the I$_h$ cluster with
$r$=4.84 a.u.. The DOS were obtained by a Lorentzian extension of the
discrete energy levels and a summation over them. The broadening
width parameter was chosen to be 0.4 eV. From Fig. 4, we can see that
the central atom contributes mainly the DOS in the bottom region of
the valence band, and the DOS around E$_F$ are contributed mostly
by the surface atoms. The total DOS in the valence-band region are
dominantly of $4d$ character,
but the compositions of $5s$ and $5p$ can be easily seen from the figures.
E$_F$, which is -4.3 eV, is found to lie just at the peak of the
minority-spin DOS. The valence band width is obtained to be 5.6 eV,
1.1 eV larger than
that of Reddy {\it et al.},\cite{reddy}
showing again the effect of enlarging
the basis set. The exchange splitting is estimated to be 0.7 eV, compared
to 0.9 eV of Reddy {\it et al.}.

In conclusion, we have presented the
electronic structure of 13-atom Rh clusters, and discussed their
anomalous magnetic properties. The ground state is found to
exhibit the icosahedral structure, and has total magnetic moment of
15$\mu$B.
This value is anomalously smaller than those
for clusters with lower symmetries, but in agreement with recent experiments.
The magnetic interactions between the central and surface atoms of the cluster
are {\it not} fully FM, and a small amount of AFM
interactions is found to be mixed in. The anomalous magnetic properties are
explained in terms of an energy parameter, which is found to be also
useful for judging whether the broadening technique is correctly used and
whether multiple input potentials must be used to reach the actual ground
state in the LSD calculations.

One of us (Y.J) gratefully acknowledges the
hospitality of GNSM in Padova, the encouragement from GCMT
in Hefei, and the financial support of CNR (Italy) and CAS
(China). The portion of the
work performed in China was supported in part by the National Natural
Science Foundation of China.

\begin{figure}
\caption{Binding energies of the I$_h$ (solid line), O$_h$ (dot-dashed line),
and D$_{3h}$ (dashed line) Rh$_{13}$ clusters vs the distance $r$ between
the central and surface atoms.}
\label{fig1}
\end{figure}

\begin{figure}
\caption{The cluster moment as a function of $r$ for the I$_h$ Rh$_{13}$
cluster
(solid line),
and for the O$_h$ and D$_{3h}$ Rh$_{13}$ clusters(dot-dashed line).}
\label{fig2}
\end{figure}

\begin{figure}
\caption{Spin-density distribution of the I$_h$ Rh$_{13}$ cluster with
$r$=4.84 a.u..
(a) is plotted in the plane passing through the five surface atoms,
(b) in the plane passing through the central and
four surface atoms.
Positive, zero and negative values of the spin-density are
indicated by full, dotted and dashed lines, respectively.}
\label{fig3}
\end{figure}

\begin{figure}
\caption{DOS for the I$_h$ Rh$_{13}$ cluster with $r$=4.84 a.u.: (a).
majority spin and (b). minority spin. Rh(1) and Rh(2) denote the central
and surface atoms respectively.}
\label{fig4}
\end{figure}

\begin{table}
\caption{The equilibrium bondlengths and binding
energies for Rh$_{13}$ clusters($r$: the distance between the center and
surface
atoms).}
\label{table1}
\begin{tabular}{ccccc}
Symmetry&\multicolumn{2}{c} {$r$ (a.u.)} &\multicolumn{2}{c}{E$_b$ (eV)}\\
&Our work&Reddy {\it et al.}\cite{reddy}&Our work
&Reddy {\it et al.}\cite{reddy} \\ \hline
I$_h$&4.84&4.84&51.16&42.6\\
O$_h$&4.95&4.90&49.81&41.3\\
D$_{3h}$&4.96&&50.71&\\
\end{tabular}
\end{table}

\begin{table}
\caption{The local and total magnetic moments($\mu_B$)of Rh$_{13}$
clusters at the equilibrium configurations.}
\label{table2}
\begin{tabular}{cccccc}
Symmetry&\multicolumn{4}{c} {Local moment}&Total \\
&Orbital&Center atom&\multicolumn{2}{c}{surface atom}&moment\\ \hline
I$_h$&$4d$&1.323&\multicolumn{2}{c} {0.993}&15\\
     &$5s$&-0.004&\multicolumn{2}{c} {0.095}&\\
     &$5p$&-0.195&\multicolumn{2}{c} {0.068}&\\
     &total&1.123&\multicolumn{2}{c} {1.156}&\\
O$_h$&$4d$&1.284&\multicolumn{2}{c} {1.313}&19\\
     &$5s$&-0.015&\multicolumn{2}{c} {0.105}&\\
     &$5p$&-0.230&\multicolumn{2}{c} {0.079}&\\
     &total&1.038&\multicolumn{2}{c} {1.497}&\\
D$_{3h}$&$4d$&1.283&1.332&1.332&19\\
        &$5s$&-0.014&0.098&0.095&\\
        &$5p$&-0.207&0.051&0.092&\\
     &total&1.062&1.481&1.509&\\
\end{tabular}
\end{table}


\begin{references}

\bibitem{chicago} For a review see, Proceedings of the Sixth International
Meeting on Small Particles and Inorganic Clusters [Z. Phys. D
 {\bf 26}, Nos. 1-4(1993)].
\bibitem{yang} Yang Jinlong, Xiao Chuanyun, Xia Shangda, and Wang Kelin,
Phys. Rev. B {\bf 48}, 12155(1993).
\bibitem{dunlap} B.I. Dunlap, Phys. Rev. A {\bf 41}, 5691(1990).
\bibitem{khanna} S.N. Khanna, S. Linderoth, Phys. Rev. Lett {\bf 67},
742(1991).
\bibitem{li} Z.Q. Li and B.L. Gu, Phys. Rev. B {\bf 47}, 13611(1993).
\bibitem{deheer} W.A. deHeer, P. Milani, and A. Chatelain,
Phys. Rev. Lett {\bf 65}, 488(1990).
\bibitem{bucher} J.P. Bucher, D.C. Douglass, and L.A. Bloomfield,
Phys. Rev. Lett {\bf 66}, 3052(1991).
\bibitem{louderback} J.G. Louderback, A.J. Cox, L.J. Lising,
D.C. Douglass, L.A. Bloomfield, Z. Phys. D {\bf 26}, 301(1993).
\bibitem{liu} F. Liu, S.N. Khanna, and P. Jena, Phys. Rev. B {\bf 43},
 8179(1991).
\bibitem{pastor} G.M. Pastor, J. Dorantes-Davila, and K.H. Benneman,
Phys. Rev. B {\bf 40}, 7642(1989).
\bibitem{douglass} D.C. Douglass, J.P. Bucher, and L.A. Bloomfield,
Phys. Rev. B {\bf 45}, 6341(1992).
\bibitem{cox} A.J. Cox, J.G. Louderback, and L.A. Bloomfield,
Phys. Rev. lett. {\bf 71}, 923(1993).
\bibitem{reddy} B.V. Reddy, S.N. Khanna, and B.I. Dunlap, Phys. Rev. Lett.
{\bf 70}, 3323(1993).
\bibitem{falicov} L.M. Falicov, Phys. World {\bf 6}(12), 24(1993).
\bibitem{delley1} B. Delley and D.E. Ellis, J. Chem. Phys. {\bf 76},
1949(1982).
\bibitem{delley2} B. Delley, D.E. Ellis, A.J. Freeman, E.J. Baerends,
and D. Post, Phys. Rev. B {\bf 27}, 2132(1983).
\bibitem{rosen} A. Rosen, D.E. Ellis, H. Adachi, and F.W. Averill,
J. Chem. Phys. {\bf 65}, 3629(1976).
\bibitem{barth} U. von Barth and L. Hedin, J. Phys. C {\bf 5}, 1629(1972).
\bibitem{perdew} J.P. Perdew and A. Zunger, Phys. Rev. B {\bf 23}, 5048(1981).
\bibitem{apai} G. Apai, J.F. Hamilton, J. Stohr, and
A. Thompson, Phys. Rev. Lett {\bf 43}, 165(1979).

\end{references}
\end{document}